\documentclass[aps,prl,preprint]{revtex4}
\usepackage{amsmath}
\usepackage{amssymb}
\usepackage{psfrag}
\usepackage{bm}
\usepackage{graphicx}
\newcommand{\D}{\mbox{\boldmath ${\cal D}$}}
\newcommand{\bee}{\begin{eqnarray}}
\newcommand{\eee}{\end{eqnarray}}
\newcommand{\ba}{\begin{array}}
\newcommand{\ea}{\end{array}}
\newcommand{\x}{\mbox{\boldmath $\zeta$}}
\newcommand{\uu}{\hat{\bf u}}

\newcommand{\La}{\Lambda^h}
\newcommand{\mx}{{x}}

\newcommand{\ka}{\kappa \bar{a}}

\newcommand{\nnabla}{\mbox{\boldmath $\nabla$}}

\begin{document}
\title{Self-diffusion of a sphere in an effective medium of rods}
\author{Bogdan Cichocki$^1$}
\email{cichocki@fuw.edu.pl}
\author{Maria L. Ekiel-Je\.zewska$^2$}
\email{mekiel@ippt.gov.pl}
\affiliation{%
$^1$  Institute of Theoretical Physics, University of Warsaw, ul. Ho\.za 69, 00-681 Warsaw, Poland,}
\affiliation{%
$^2$  Institute of Fundamental Technological Research,
Polish Academy of Sciences,
ul. \'Swi\c etokrzyska 21,
00-049 Warsaw, 
Poland
}%

\date{\today}

\begin{abstract}
Self-diffusion of a sphere in a network of rods is analyzed theoretically. Hydrodynamic interactions are taken into account according to the model of Dhont et al., under the assumption that $\ka\! << \!1$ and $\bar{a}/L\!<<\!1$, where $1/\kappa$ is the network hydrodynamic screening length, $\bar{a}=a+D/2$, and $a$ is the sphere radius, while $D$ and $L$ are the diameter and length of a rod, respectively. Simple expressions for the diffusion coefficients are derived and shown to be independent of $L$.
\end{abstract}

\maketitle
Tracer diffusion of a spherical particle in a suspension of slender particles is a common process in 
various biological systems, such as proteins in 
suspensions of F-actins, nucleosome core particles in DNA dispersions or colloidal particles  in polymer systems. 

In Refs.~\cite{Dhont1,Dhont2,Dhont3}, this process has been used to determine the structure of such networks, in isotropic and nematic phases.
A theoretical model of the screened hydrodynamic interactions between a tracer sphere and motionless rods has been constructed, by adding an effective friction force to the hydrodynamic equations. The model was  
used to deduce from the experiment the screening length of a network of fd viruses as a  function of the rod concentration. 
In this communication, we follow the model and derive simple expressions for the self-diffusion tensor, helpful 
in further studies of rod networks.

According to the fluctuation-dissipation theorem, the self-diffusion tensor $\D$
is related to the translational-translational hydrodynamic friction tensor $\x$,
\bee
\D=k_BT \langle \x \rangle^{-1},\label{fd}
\eee
where $\langle ... \rangle$ denotes the average over positions and orientations of the rods. 
In the above equation, the axial symmetry of the system has been taken into account. In this case, the translational-rotational hydrodynamic friction tensor vanishes. Therefore the external force 
$\langle {\bf F}_{ext} \rangle$ needed to move the sphere with a 
velocity ${\bf v}_p$ does not generate a sphere rotation, and it 
is given by the relation,
\bee
\langle{\bf F}_{ext} \rangle &=& \langle \x\rangle \cdot {\bf v}_p.
\eee

In Refs.~\cite{Dhont1,Dhont2,Dhont3}, the medium velocity ${\bf v}$ and pressure $p$ are described by the Debye-B\"uche-Brinkman equation~\cite{DebyeBueche,Brinkman}, 
\bee
\eta(\nnabla^2 {\bf v} -\kappa^2 {\bf v})-\nnabla p = {\bf 0}, \hspace{1cm} \nnabla \cdot {\bf v}=0,\label{DBB}
\eee
with the hydrodynamic screening length $1/\kappa$, and the isotropic Green tensor in both isotropic and nematic phases,
\bee
{\bf T}_{s}({\bf r}) &=& \frac{1}{4\pi \eta r}[h_1(x){\bf I} +h_2(x)\hat{\bf r} 
\hat{\bf r}],\label{Tiso}
\eee
where $r=|{\bf r}|$, $\hat{\bf r}={\bf r}/r$, $x=\kappa r$ and 
\bee
h_1(x) &=& -\frac{1}{x^2} +\left(1+\frac{1}{x} +\frac{1}{x^2}\right)\exp(-x), \label{h1}
\eee
\bee
h_2(x) &=& \frac{3}{x^2} -\left(1+\frac{3}{x} +\frac{3}{x^2}\right)\exp(-x). \label{h2} 
\eee
The friction and diffusion tensors for a single sphere of radius $a$ are isotropic,
$\x_{0}\!=\!\zeta_0 {\bf I}$, $\D_0 \!=\!{\cal D}_0 {\bf I}$, with
$\zeta_0\!= \!6\pi \eta a (1\! +\! \kappa a \!+\!(\kappa a)^2/3)$~\cite{Brinkman} and 
${\cal D}_{0}\!=\!k_bT/ \zeta_0$. In Refs.~\cite{Dhont1,Dhont2,Dhont3}, $\kappa a<< 1$, and 
$\zeta_0\!\approx \!6\pi \eta a$. 

In the network of $N$ rods, the cluster expansion of the friction tensor is performed. After the averaging,
\bee
\frac{\langle \x \rangle}{\zeta_0} =   {\bf I} + \sum_{j=1}^{N} \langle {\bf M}({\bf R}_j,\hat{\bf u}_j) \rangle + ....\label{withHI}
\eee
where ${\bf M}({\bf R}_j,\hat{\bf u}_j)$  is the correction from hydrodynamic interactions between the sphere, located at the center of the reference frame, and single rods, labeled $j$, with their centers located at ${\bf R}_j$ and oriented along $\hat{\bf u}_j$.

For the isotropic phase, $\D={\cal D}_{iso} {\bf I}$. For the nematic phase, 
with the axial direction ${\bf e}_{\parallel}$, the diffusion tensor 
${\D} = {\cal D}_{\parallel}{\bf e}_{\parallel} {\bf e}_{\parallel} + {\cal D}_{\perp}({\bf I} - {\bf e}_{\parallel} {\bf e}_{\parallel})$.
With the use of volume fraction $\phi=\bar{\rho}\pi LD^2 /4$ (where $\bar{\rho}$ is the number density of rods, D is their thickness and L is their length) and Eqs.~\eqref{fd}, \eqref{withHI}, the diffusion coefficients ${\cal D}\!=\!{\cal D}_{iso},\, {\cal D}_{\parallel},\, {\cal D}_{\perp}$ are rewritten~as,
\bee
{\cal D}/{\cal D}_{0} &=& (1+ \alpha \phi)^{-1},
\eee
with 
$\alpha \!=\! \alpha_{iso},\,\alpha_{\parallel},\,\alpha_{\perp}$, respectively, where 
\bee
\hspace{-0.6cm}\alpha_{iso} \!\!\!&=& \!\!\!\frac{4}{3\pi D^2 L} 
\int \!\!  d{\bf R}\;\overline{\exp[-\beta V({\bf R},\hat{\bf u})] \mbox{Tr} {\bf M}({\bf R},\hat{\bf u})},\label{i}
\eee
$\alpha_{\parallel}$ is obtained when $\mbox{Tr} {\bf M}/3$ in Eq.~\eqref{i} is replaced by 
${\bf e}_{\parallel} \cdot {\bf M}\cdot {\bf e}_{\parallel}$, and  
the coefficient $\alpha_{\perp}$ is evaluated from the relation
$
3\alpha_{iso} = (\alpha_{\parallel}+2\alpha_{\perp}).\label{para2}
$
In Eq. \eqref{i}, the explicit form of the average $\langle ... \rangle$ has been specified, with a direct sphere-rod interaction potential $V({\bf R},\hat{\bf u})$, and
\bee
\overline{f(\uu)} \equiv \int d\hat{\bf u}\; f(\uu) \; P_0(\hat{\bf u}|S),\label{avan}
\eee
where  $P_0(\hat{\bf u}|S)$ is an axially symmetric probability distribution function, which depends on 
$\hat{\bf u}$ only through the angle $\theta$, defined by 
$
\cos \theta = \hat{\bf u}\cdot {\bf e}_{\parallel}%. \label{theta}
$.  
Here $S$ denotes 
the orientational order parameter, 
\bee
S = \frac{3}{2} \overline{\cos^2 \theta}  -\frac{1}{2}.\label{S}
\eee

To evaluate ${\bf M}$ in Eq.~\eqref{withHI}, the rods are treated as 
motionless, and the bead model of each rod 
is assumed, with the odd number of identical beads $K\!=\!2n\!+\!1\!=\!L/D$, and their diameter $D$.  
For $a/L\! <\!< \!1$ and $K\! >\!> \!1$, it is useful to represent ${\bf M}$ as the multiple scattering series, with the scattering sequences starting and finishing at the 
spherical particle; the leading contribution to ${\bf M}$ involves only a single scattering from the rod. This scattering involves flows incoming onto each bead $-n \!\le \!i\!\le\! n$, located at ${\bf r}_i\!=\!{\bf R}\!+\!iD\hat{\bf u}$, and the flows outgoing from any bead $-n \!\le \!j\!\le\! n$, located at ${\bf r}_j\!=\!{\bf R}\!+\!jD\hat{\bf u}$,
\bee
{\bf M}({\bf R},\hat{\bf u})\!\!\! &=& \!\!\!6\pi\eta a \!\!\sum_{i=-n}^{n} \sum_{j=-n}^{n}\!\! {\bf T}_s({\bf r}_i) \cdot \x_{ij}(\hat{\bf u}) \cdot {\bf T}_s({\bf r}_j).\hspace{0.5cm}\label{MM}
\eee
In the above equation, the hydrodynamic interactions between the rod beads $i$ and $j$ are described by the friction tensors $\x_{ij}(\hat{\bf u})$, which in general determine the external force ${\bf F}_j$ on a particle $j$ needed to cause the translational velocities ${\bf U}_i$ of all the particles $i$,
\bee
{\bf F}_i &=& \sum_{j} \x_{ij}(\hat{\bf u}) \cdot {\bf U}_j.\label{defzetaij}
\eee
The hydrodynamic interactions between the rod beads are described by the point-particle model.
Following Ref.~\cite{Dhont3}, 
we approximate Eq.~\eqref{MM} by taking the limit of $n\rightarrow \infty$, $\kappa D\rightarrow 0$, with $L=\mbox{const}$,
\bee
{\bf M}({\bf R},\hat{\bf u}) \!\!\!&=& \!\!\!- \,\frac{24(\pi\eta)^2 a}{\ln (\kappa D)}\!
\int_{-L/2}^{L/2}\!\!\!\! dl \,{\bf T}_s({\bf R}') \cdot 
\left( \!{\bf I}\!-\!\frac{\hat{\bf u} \hat{\bf u}}{2}\!\right) \!
\cdot {\bf T}_s({\bf R}'),\nonumber\\
\label{Msamo}
\eee
where ${\bf R}'={\bf R} + l\hat{\bf u}$.

In Refs.~\cite{Dhont1,Dhont2,Dhont3}, the 
hard-core interaction potential is assumed. In this case, 
\bee
\exp[-\beta V({\bf R},\hat{\bf u})] &=& 1-\chi({\bf R},\hat{\bf u}),\label{hc}
\eee
where $\chi({\bf R},\hat{\bf u})$ is the characteristic function equal to one inside the cylinder centered at ${\bf R}$, oriented along $\hat{\bf u}$, of length $L$ and radius  $\bar{a}=a+D/2$, where 
 we used $\bar{a}/L<< 1$ to approximate the shape of the rod close to its ends.

Eqs.~\eqref{i} and \eqref{Msamo} allow to express the hydrodynamic coefficients in terms of
four-dimensional integrals. In Refs.~\cite{Dhont1,Dhont2,Dhont3}, these integrals have been carried out numerically for a specific value of $\Gamma = L/\bar{a}$. In this paper, we will carry out explicitly three of these integrations, giving much simpler expressions for the 
hydrodynamic coefficients. We will also derive simple asymptotic expressions in the limit $\ka \rightarrow 0$.

In the following, we explain our method focusing on the derivation of $\alpha_{iso}$. The other hydrodynamic coefficients are obtained by a straightforward generalization.
From Eqs.~\eqref{i} and \eqref{Msamo} it follows that  we need
to calculate 
\bee
\hspace{-0.3cm}
\mbox{Tr}\! \left[{\bf T}_s({\bf R}')\cdot\left( \!{\bf I}\!-\!\frac{\hat{\bf u} \hat{\bf u}}{2}\!\right)   \cdot {\bf T}_s({\bf R}')\right]\!\!
\!\!\! &=&\!\!\frac{\kappa^2}{8(4\pi\eta)^2} A(\mx,s),\hspace{0.4cm}\label{trA}
\eee
which is a
scalar function of $\mx\!=\!\kappa R'$ and $s\!=\!\hat{\bf u} \!\cdot \!{\bf R}'/R'$\!, with
\bee
A(\mx,s)\!\!&=&\!\!-\,[ b_0(\mx)+s^2 b_2(\mx)]/\mx^2,
\eee
and
\bee
b_0(\mx)\!\!&=&\!\! -\;4[5h_1^2(\mx)+4h_1(\mx)h_2(\mx)+2h_2(\mx)^2],\hspace{0.4cm}\label{A1}\\
b_2(\mx)\!\!&=&\!\! 4[2h_1(\mx)+h_2(\mx)]h_2(\mx).\label{A3}
\eee
For a fixed $\uu$, we change the integration variables in Eq.~\eqref{i}, ${\bf R}\!\rightarrow \!{\bf R}'\!=\!{\bf R}\!+\!l\hat{\bf u}$. With Eqs.~\eqref{Msamo}-\eqref{trA}, we~obtain,
\bee
\alpha_{iso} &=& -\;\dfrac{\kappa a}{(\kappa D)^2\ln(\kappa D)} \La_{iso},\label{scala}
\eee
where the hydrodynamic scaling function has the form,
\bee
\La_{iso} 
\!\!\! &=&\!\!\!\dfrac{\kappa^3}{4\pi L}\int_{-L/2}^{L/2}\!\!\!\!dl\; \int\!\! d{\bf R}'\, [1-\chi({\bf R},\hat{\bf u})]A(\mx,s).\hspace{0.7cm}
\label{H-M}
\eee
The averaging with respect to the rod orientation has been omitted, because in the isotropic phase, the integration with respect to ${\bf R}'$ results in an expression independent of $\uu$.

The hydrodynamic scaling function \eqref{H-M} is decomposed into uncorrelated and correlated parts, corresponding to the integrands proportional to $1$ and $(-\chi)$, respectively,
\bee
\La_{iso} &=&\La_{iso,0} + \La_{iso,c}.
\eee

Calculating $\La_{iso,0}$, we change ${\bf R}'$ into 
spherical integration variables $\mx$, $s$ and~$\phi$. We obtain,
\bee
\La_{iso,0} \!\!&=& \!\!-\; \int_0^{\infty} d\mx \;%\int_0^1 ds \;
\left[ b_0(\mx)+\frac{1}{3}b_2(\mx)\right]\!.\hspace{0.6cm}
\eee
The above integral is evaluated analytically by substituting the explicit expressions~\eqref{A1}-\eqref{A3}, with $h_1$ and $h_2$ given by Eqs.~\eqref{h1}-\eqref{h2}, 
and taking the limit $\epsilon \rightarrow 0$ of 
the exponential integrals $E_n(\epsilon)$, $E_n(2\epsilon)$. 
Finally,
\bee
\La_{iso,0} \!\!\!&=& \!\!\!
\frac{20}{3}.\label{unc}
\eee

The correlated part of the hydrodynamic scaling function follows from Eq.~\eqref{H-M}, if
cylindrical coordinates $z',\;\rho,\;\phi$ are introduced to represent ${\bf R}'$, with the $z'$ axis along $\hat{\bf u}$. In addition, we take into account that 
the integrand does not depend on $\phi$, and therefore, 
\bee
\La_{iso,c} \!\!&=&\!\!   -\;\dfrac{\kappa^3}{2 L}\int_{-L/2}^{L/2} \! \!\! d{l} \int_{l-L/2}^{l+L/2} \!\!  dz' \int_0^{\bar{a}} \!\! \rho\, d\rho\, A(\mx,s).
\hspace{0.7cm}\label{AA}
\eee
We now change the order of the integration with respect to $l$ and $z'$ in Eq.~\eqref{AA}.
The integrand does not depend on $l$ and is symmetric under $z'\!\rightarrow\! -z'$. 
Therefore 
\bee
\La_{iso,c} \!\!&=&\!\!   -\;\dfrac{\kappa^3}{L} \int_{0}^{L} \!\!  dz' (L-z')\;\int_0^{\bar{a}} \!\! \rho d\rho \;
A(\mx,s),
\label{H2}
\eee
and with the change of variables, $z',\rho \rightarrow \mx,s$, 
\bee
\La_{iso,c} 
\!\!\!&=&\!\!\! - \!\left[\! \int_0^{\kappa \bar{a}} \!\!\!\!\!\!\!\!d\mx \!\int_0^1\!\!\!\!\! ds 
+ \!\!\int_{\kappa \bar{a}}^{\kappa L} \!\!\!\!\!\! \!\! d\mx
\!\int_{\sqrt{1-(\kappa \bar{a}/{\mx})^2}}^1\!\! ds 
\right]\!\!\left(\!1\!\!-\!\frac{\mx s}{\kappa L}\right) \!\mx^2\!A(\mx,s)\!.\nonumber\\ \label{two}
\eee
The integrand in Eq.~\eqref{two} is a polynomial in $s$, therefore the integration with respect to $s$ is 
straightforward and leads to the functions,
\bee
\!\!\!
p_n(y)\!\!\! &=&\!\!\! (n\!+\!1) \int_{\sqrt{1-y^2}}^1 ds\; s^n \!= \!
\left[ 1-\left(1\!-\!{y^2} \right)^{(n+1)/2}\right]\!\!.\hspace{0.7cm}\label{pn}
\eee
Combining Eqs.~\eqref{unc} and \eqref{two}, 
we express the hydrodynamic scaling function by single integrals only, 
\bee
\La_{iso}\!\!\! &=& \frac{20}{3} + (I_1+I_2)-\frac{1}{\kappa L} (J_0+J_2),\label{Laiso}
\eee
where
\bee
I_n \!\!&=&\!\!\!\frac{1}{n+1}\left[ \int_0^{\epsilon}\!\!d\mx +\int_{\epsilon}^{\kappa L} \!\!d\mx\; p_n\!\left(\frac{\epsilon}{\mx}\right)\right] \, b_n(\mx),\label{In}\\
J_n \!&=& \!\frac{1}{n+2}\left[ \int_0^{\epsilon}\!\!d\mx+\int_{\epsilon}^{\kappa L}\!\! d\mx\; p_{n+1}\left(\frac{\epsilon}{\mx}\right)\right] \mx b_n(\mx),\hspace{0.6cm}\label{Jn}
\eee
with $\epsilon=\ka$ and $n=0,2$. The above integrals can be 
evaluated numerically with a high precision. 

Assuming a fixed value of $\kappa L$, we now evaluate 
asymptotic form of  $\La_{iso}$ 
for $\epsilon\!\rightarrow 0$. 
Below we derive the asymptotic expressions for $I_0$; all the other terms 
are obtained by the analogical reasoning. 
To make the presentation more clear, we denote $f(\mx)=b_0(\mx)$. 
First, using the expansion,
\bee
f(\mx) &=& f^{(0)} + f^{(1)}\mx + o(\mx), \hspace{1cm} \mbox{for } \mx\rightarrow 0.
\hspace{0.5cm}\label{og}
\eee
we represent $I_0$ as a sum of two integrals, with $f(x)$ replaced by $f^{(0)}\! +\! f^{(1)}\mx$ and $f(x)\!-\!f^{(0)}\! -\! f^{(1)}\mx$, respectively.
The first integral is carried out explicitly and then the dominant terms 
for small $\epsilon$ are kept. Asymptotic behavior of the second integral 
follows from the expansion $
p_n\left( y\right) \!=\! (n\!+\!1)y^2/2\! +\! o(y^2)$, valid for $y<1$.
Combining both integrals, we obtain,
\bee
I_0 \!\!&=&\!\! \epsilon\,\frac{\pi}{2} f^{(0)} +\frac{\epsilon^2}{2} \left[
C_{\{f\}}(\kappa L)
 + \frac{f^{(1)}}{2} - f^{(1)} 
\ln \frac{\epsilon}{2} \right] + o(\epsilon^2),\nonumber \\
\eee
where
\bee
C_{\{f\}}(\kappa L)\!\! &=&\!\! \!\int_0^{\kappa L} \!\! \!\!\!d\mx\; \frac{f(\mx)\!-\!f^{(0)}\! -\! f^{(1)}\mx}{\mx^2}-\frac{f^{(0)}}{\kappa L} 
+f^{(1)} \ln{\kappa L}.\nonumber \\ \label{dC}
\eee
Asymptotics of the other integrals is also evaluated. $I_2$ has the same form 
as $I_0$, but with different numerical coefficients of the first and the third 
terms. $J_n$ do not contain a term linear in $\epsilon$. Therefore, 
as expected, in the limit $\epsilon \rightarrow 0$, the term $(J_0+J_2)/(\kappa L)$ is 
negligible in comparison to $(I_0+I_2)$, which scales as
\bee
\La_{iso,c} \!\!&=&\!\! -\;\epsilon \frac{19\pi}{4} + \frac{\epsilon^2}{2}
\left[ C_{\{b_0+b_2\}}(\kappa L)
 +\frac{32}{3} -16\ln\frac{\epsilon}{2} \right]\!,
\nonumber \\
\label{maiso}
\eee
In deriving the above formula, 
the expansion of the
Green functions \eqref{h1} and \eqref{h2}, $h_1(\mx)=1/2 -2 \mx/3 + o(\mx)$ and $h_2(\mx)=1/2  + o(\mx)$, has been used to find explicitly the expression \eqref{og}.
Then, the function
$C_{\{b_0\!+\!b_2\}}(\kappa L)$ is approximated 
by its value at infinity, 
evaluated numerically,
\bee
C_{\{b_0\!+\!b_2\}}(\mbox{\small${+\infty}$})
\!\!\!&=&\!\!\! 
-4.944.\label{valueC}
\eee
For large $\kappa L$, we 
analyze 
the difference,
\bee
C_{\{f\}}(\kappa L)-C_{\{f\}}
(\mbox{\small${+\infty}$})
\!\!&=&\!\!-\int_{\kappa L}^{\infty} \!d\mx\,\frac{f(\mx)}{\mx^2}.\label{di}
\eee
With the increasing $\mx$, the integrand decays to zero, with terms proportional to $1/\mx^6$, $\exp(-\mx)/\mx^4$, $\exp(-2\mx)/\mx^2$. For $\kappa L \gtrsim 3$, the difference \eqref{di} is negligible. On the other hand, for $\kappa L<<1$,
\bee
 \frac{\epsilon^2}{2}C_{\{f\}}(\kappa L) \approx - \frac{\epsilon^2}{2\kappa L} f^{(0)}.
\eee
For $f^{(0)}=b_0^{(0)}+b_2^{(0)}=-8$, and the above term gives a significantly smaller contribution to $\La_{iso,c}$ than $19\pi\epsilon /4$, because $\epsilon/\kappa L << 1$. 

Adding the uncorrelated contribution, $\La_{iso,0}=20/3$, to the correlated one,~\eqref{maiso} with \eqref{valueC}, we obtain a simple approximate expression for the hydrodynamic scaling function in the isotropic phase,
\bee
\La_{iso} \!\!&=&\!\! \frac{20}{3} -\;\frac{19\pi}{4}\epsilon +[8.41-8\ln \epsilon]\epsilon^2,\label{mainiso}
\eee
valid for $\epsilon=\ka<<1$ and $\bar{a}/L<<1$.

We have applied the same reasoning to evaluate also
the hydrodynamic scaling functions $\alpha_{\parallel,\perp}$, introduced in Refs.~\cite{Dhont1,Dhont2,Dhont3} for the nematic phase.  
They are expressed in terms of $\La_{\parallel,\perp}$
by equations analogical to~\eqref{scala}, with
\bee
\La_{\parallel}\!\!&=&\!\!\La_{iso}-2S\La_s,\\
\La_{\perp}\!\!&=&\!\!\La_{iso}+S\La_s,
\eee
where $S$ is defined in Eq. \eqref{S} and $\La_s$ is a function of $\epsilon=\ka$, given by the approximate expression,
\bee
\La_{s} \!\!&=&\!\!\frac{14}{15} -\;\frac{13\pi}{32}\epsilon + 1.024\epsilon^2. \label{mains}
\eee

\begin{figure}[ht]
\begin{center}
\psfrag{          kbara}{$\kappa \bar{a}$}
\includegraphics[width=8.5cm]{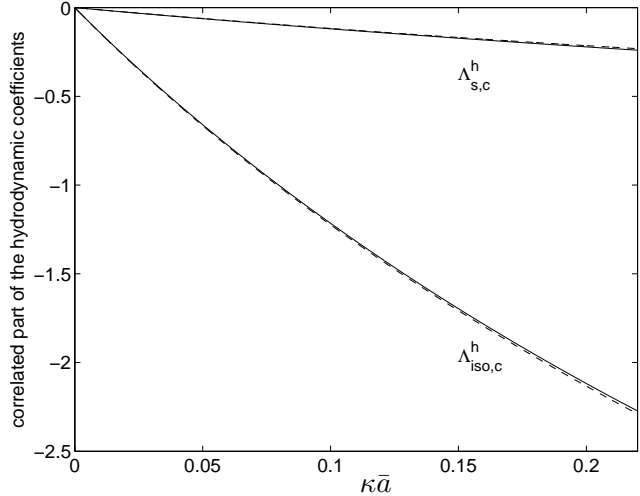}
\caption{Hydrodynamic scaling functions $\La_{iso,c}$ and $\La_{s,c}$ versus $\kappa  \bar{a}$. 
Approximate expressions \eqref{mainiso} and \eqref{mains} (dashed lines) and results obtained from exact formulas ~\eqref{Laiso}-\eqref{Jn} and their analogs for the nematic phase (solid lines).
}\label{fi2}
\end{center}
\end{figure}

The correlated parts of the approximate expressions \eqref{mainiso} and \eqref{mains} (independent of $\Gamma \equiv L/\bar{a}$) are plotted in Fig.~\ref{fi2}, together with the accurate results of numerical integration of Eqs.~\eqref{Laiso}-\eqref{Jn} and their analogs for the nematic phase (performed for $\Gamma =90.7$). 
For $\ka \lesssim 0.2$, precision of the approximation for $\La_{s}$ is  better than 1\%, and for $\La_{iso}$ better than 0.3\%. 

\begin{figure}[ht]
\begin{center}
\includegraphics[width=8.6cm]{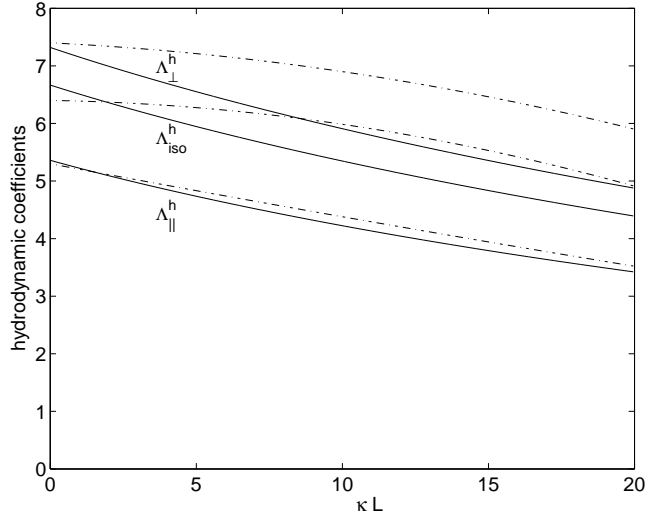}
\caption{Hydrodynamic scaling functions  $\La_{\perp}$, $\La_{iso}$, $\La_{\parallel}$ (top-down).
Solid lines: results calculated from the exact expressions,~\eqref{Laiso} and its nematic analog; dashed-dot lines: numerical results from 
Ref.~\cite{Dhont3}.}\label{fi1}
\end{center}
\end{figure}

The specific value of $\Gamma =90.7$
is chosen to match the experiments from 
Ref.~\cite{Dhont3}.
To compare with the results of Ref.~\cite{Dhont3}, we evaluate the hydrodynamic scaling functions from 
the exact expressions, \eqref{Laiso} and its analog for the nematic phase, and plot the results in Fig.~\ref{fi1}, 
together with the numerical results of Ref.~\cite{Dhont3}, obtained for $S=0.7$ and shown in the left plot of their Fig.~3.
The pronounced differences between their and our curves are probably caused by 
inaccuracy of the four-dimensional numerical integration performed in 
Ref.~\cite{Dhont3}. In our approach, three integrals are carried out 
analytically, what allows to improve significantly precision of the 
numerical results. In addition, we performed asymptotic expansion of 
the single integral which is left, and proposed simple expressions \eqref{mainiso}-\eqref{mains} 
for $\La_{iso,s}$, 
valid in the whole range of applicability of the 
physical model used to describe the system, that is for $\ka<<1$ and $\bar{a}/L<<1$. 

Our important conclusion is that in practice, $\La_{iso,s}$
depend on $\ka$ only.  This result 
indicates that the proper argument of the scaling functions is $\ka$ 
rather than $\kappa L$ used in 
Ref.~\cite{Dhont3}.
 
Possible applications are the following. In Refs.~\cite{Dhont1,Dhont2,Dhont3}, diffusion of spheres in isotropic and nematic networks of rods 
has been modelled, based on the isotropic Debye-Bueche-Brinkman equation. The model allows to 
use the measured values of the self-diffusion coefficients to 
determine the hydrodynamic screening length $1/\kappa$ of the network of rods. 
The simple and accurate expressions for the hydrodynamic coefficients derived in 
this paper can be used to make this procedure efficient.

In future, predictions of the model from Ref.~\cite{Dhont3} can be verified experimentally by 
measuring self-diffusion of spheres with different radii $a$ in the same network of the rods,
and checking if $\kappa$  does not depend on $a$.  In the nematic phase, 
the network of rods should in general be described by the non-isotropic Debye-Bueche-Brinkman equation. In the following study, the calculation of the hydrodynamic coefficients described in this paper will be modified for such a non-isotropic medium. 

\acknowledgements
We thank J. K. G. Dhont and K. Kang for helpful discussions.


\begin{thebibliography}{99}
\bibitem{Dhont1}
K. Kang, J. Gapi\'nski, M. P. Lettinga, J. Buitenhuis, G. Meier, M. Ratajczyk, J. K. G. Dhont and A. Patkowski, J. Chem. Phys. 
{\bf 122}, 044905 (2005).
\bibitem{Dhont2}
K. Kang, A. Wilk, J. Buitenhuis, A. Patkowski, and J. K. G. Dhont, J. Chem. Phys. 
{\bf 124}, 044907 (2006).
\bibitem{Dhont3}
K. Kang, A. Wilk, A. Patkowski, and J. K. G. Dhont, J. Chem. Phys. {\bf 126}, 214501 (2007).
\bibitem{DebyeBueche}
P. Debye and A. M. Bueche, J. Chem. Phys. {\bf 16}, 573 (1948).
\bibitem{Brinkman}
H. C. Brinkman, Appl. Sci. Res. {\bf A1}, 27 (1947).
\bibitem{Guzowski}
J. Guzowski, B. Cichocki, E. Wajnryb, and G. C. Abade, J. Chem. Phys. {\bf 128}, 094502  (2008). 
\end{thebibliography}
\end{document}